\documentclass[9pt,twocolumn,twoside,a4paper]{pnas-new}
\templatetype{pnasresearcharticle}
\usepackage{epsfig,amsmath,amssymb,graphicx,color}
\usepackage{color}
\usepackage{soul}
\setboolean{displaywatermark}{false}

\DeclareMathOperator{\atanh}{atanh}
\def \qea {q_\text{\tiny EA}}

\title{Strong ergodicity breaking in aging of mean field spin glasses}
\author[a,1]{Massimo Bernaschi}
\author[b,1]{Alain Billoire}
\author[c,d,e,1]{Andrea Maiorano}
\author[c,e,f,1]{Giorgio Parisi}
\author[c,e,f,1,2]{Federico Ricci-Tersenghi}

\affil[a]{Istituto per le Applicazioni del Calcolo (IAC), Consiglio Nazionale delle
  Ricerche (CNR), P.le A. Moro 2, 00185 Rome, Italy }
\affil[b]{Institut de Physique Th\'eorique, Universit\'e Paris Saclay, CNRS, CEA, F-91191, Gif-sur-Yvette, France}
\affil[c]{Dipartimento di Fisica, Sapienza Universit\`a di Roma, P.le Aldo Moro 5, 00185 Rome, Italy}
\affil[d]{Insituto de Biocomputaci\'on y F\'isica de Sistemas Complejos (BIFI), Mariano Esquillor G\`omez 50018 Zaragoza, Spain}
\affil[e]{Istituto Nazionale di Fisica Nucleare, Sezione di Roma I, P.le A. Moro 5, 00185 Rome, Italy}
\affil[f]{Institute of Nanotechnology (NANOTEC) - CNR, Rome unit, P.le A. Moro 5, 00185 Rome, Italy}

\significancestatement{Understanding the relaxation dynamics on complex energy landscapes is
a key aspect not only for physical processes, but also for many
applications where the minimization of a complicated cost function is
required. Mean-field models with disorder have been and still are the
privileged playing field to try to understand such complex dynamical
behavior. We report the results of extraordinary long numerical
simulations (more than $2^{64}$ spin updates) of the prototypical
mean-field model of spin glasses. We uncover that, contrary to common
expectation, the off-equilibrium aging dynamics at low temperatures
undergoes a strong ergodicity breaking and thus asymptotically remains
trapped in a confined region of the configurational space. Our results
ask for a fundamental revision of models for the aging dynamics.}

\equalauthors{\textsuperscript{1}M.B., A.B., A.M., G.P. and F.R.T. contributed equally to this work.}
\correspondingauthor{\textsuperscript{2}To whom correspondence may be addressed. E-mail: \texttt{federico.ricci@uniroma1.it}}

\begin{abstract} 
Out of equilibrium relaxation processes show aging if they become
slower as time passes. Aging processes are ubiquitous and play a
fundamental role in the physics of glasses and spin glasses and in
other applications (e.g.\ in algorithms minimizing complex cost/loss
functions).

The theory of aging in the out of equilibrium dynamics of mean- field
spin glass models has achieved a fundamental role, thanks to the
asymptotic analytic solution found by Cugliandolo and Kurchan. However
this solution is based on assumptions (e.g.\ the weak ergodicity
breaking hypothesis) which have never been put under a strong test
until now.

In the present work we present the results of an extraordinary large
set of numerical simulations of the prototypical mean-field spin glass
models, namely the Sherrington-Kirkpatrick and the Viana-Bray models.
Thanks to a very intensive use of GPUs, we have been able to run the
latter model for more than $2^{64}$ spin updates and thus safely
extrapolate the numerical data both in the thermodynamical limit and
in the large times limit.

The measurements of the two-times correlation functions in isothermal
aging after a quench from a random initial configuration to a
temperature $T<T_c$ provides clear evidence that, at large times, such
correlations do not decay to zero as expected by assuming weak
ergodicity breaking.

We conclude that strong ergodicity breaking takes place in mean-field
spin glasses aging dynamics which, asymptotically, takes place in a
confined configurational space. Theoretical models for the aging
dynamics need to be revised accordingly.
\end{abstract}

\dates{\today}       
\doi{\url{www.pnas.org/cgi/doi/10.1073/pnas.XXXXXXXXXX}}

\begin{document}
\maketitle
\thispagestyle{firststyle}
\ifthenelse{\boolean{shortarticle}}{\ifthenelse{\boolean{singlecolumn}}{\abscontentformatted}{\abscontent}}{}

\dropcap{A}ging is a fundamental process in out of equilibrium relaxation
dynamics. It refers to the observation that relaxation or correlation
timescales grow without bound in time, so the system under study looks slower
and slower as time goes by. 
This phenomenon has been initially discovered experimentally in structural
glasses \cite{struik1976physical,struik1977physical} and spin glasses
\cite{mydosh1993spin,vincent1997slow}, but since then it has been found to be a general feature of glassy systems \cite{bouchaud1998out,amir2012relaxations}.

The importance of observing and properly describing the aging phenomena is due
to their strong connection to the energy landscape where the relaxation
dynamics takes place. Understanding the relaxation dynamics in more or
less rough energy landscapes is a very interesting and essentially still open
problem with important applications. Just to highlight a recent and very
active topic: the training of artificial neural networks --- that have
recently proven to be so effective --- is performed by minimizing the loss
function, i.e. performing a sort of relaxation dynamics in a very high-dimensional
space \cite{lecun2005loss,goodfellow2016deep}.

The dimension of the space where the dynamics takes place is indeed a crucial
aspect. While a dynamics relaxing in a low dimensional space can not be too
surprising, when the dynamics happens in a very high-dimensional space, our
intuition may easily fail in imaging the proper role of entropic and energetic
barriers and thus the description of the out of equilibrium dynamics becomes
very challenging.

Actually, any statistical mechanics model in the thermodynamic limit does 
perform a dynamics in a very high-dimensional space. In few fortunate cases
(e.g., ordered models undergoing coarsening dynamics \cite{bray2003coarsening,cugliandolo2015}) the out of
equilibrium dynamics can be well described with a reduced number of parameters.
However, in the more general case of disordered models, our understanding is
still limited and based either on numerical simulations or on the analytical
solution of restricted classes of models: mainly trap and mean field
models.

Trap models~\cite{bouchaud1992} provide a very simplified description of aging dynamics in
disordered systems by assuming that the dynamics proceeds by ``jumps'' between
randomly chosen states. This strong assumption allows for an analytic
solution, but it is not clear to what extent the actual microscopic dynamics
in a generic disordered model does satisfy such a hypothesis.

In this work we focus on disordered mean field models, in particular on the
well-known Sherrington-Kirkpatrick (SK) model~\cite{sk1975,sk1978} defined by the following
Hamiltonian
\begin{equation}
H_\text{\tiny SK} = -\sum_{i<j} J_{ij} s_i s_j\,,
\end{equation}
where the $N$ Ising spins $s_i$ interact pairwise via quenched random Gaussian
couplings $J_{ij}\sim \mathcal{N}(0,1/N)$ having zero mean and variance $1/N$.
This model is the prototype for disordered models having a continuous phase
transition from a paramagnetic phase to a phase with long range spin glass
order (in the SK model it takes place at a temperature $T_c=1$).

Thanks to the fact that couplings become very weak in the thermodynamic limit,
the off-equilibrium dynamics of mean field models can be written in terms of
two times correlation and response functions
\begin{equation*}
C(t,t') = \frac1N \sum_{i=1}^N \langle s_i(t) s_i(t')\rangle\,,\quad
R(t,t') = \frac1N \sum_{i=1}^N \frac{\partial\langle s_i(t) \rangle}{\partial h_i(t')}\,.
\end{equation*}
The angular brackets represent the average over the dynamical trajectories and an infinitesimal time-dependent field is added to the Hamiltonian as $-\sum_i h_i s_i$ to compute responses.

In the large $N$ limit $C(t,t')$ and $R(t,t')$ do satisfy a set of integro-differential equations~\cite{crisanti1993sphericalp,cugliandolo1993analytical} and the solution to these
equations provides the typical decay of correlations in a very large sample of
the SK model. Although their exact solution is unknown, Cugliandolo and
Kurchan (CK) found an ansatz that, under some hypothesis, solves the equations
in the large times limit~\cite{cugliandolo1994out}.
These equations have been rederived rigorously in some particular cases \cite{arous2006cugliandolo}.

The CK asymptotic solution has become very popular and its consequences have
been investigated in detail \cite{franz1998measuring,herisson2002fluctuation}.  It is also often used as the theoretical basis
for the analysis of numerical data \cite{franz1995fluctuation,marinari2000replica,baity2017statics}.  Notwithstanding its success, the CK
solution has never been put under a severe numerical test, due to the
difficulties in simulating very large samples of the SK model.  In particular
we are not aware of any really stringent numerical test on the assumptions
made to derive it.

One of the main hypothesis underlying the CK asymptotic solution is that one-time quantities converge to their equilibrium value.
This has been further used as a key assumption to derive the connection between statics and dynamics \cite{franz1999response}.
However, this assumption is far from obvious, given the existence of many mean-field models showing a random first order transition (RFOT) where it is apparent how that assumption may be not satisfied: the prototypical model with a RFOT is the spherical $p$-spin model where the energy relaxes to a value far from the equilibrium one if the temperature is below the dynamical transition temperature \cite{cugliandolo1993analytical}.
At variance to models with a RFOT, in spin glass models undergoing a continuous transition the common belief has been to assume convergence to equilibrium, but even these models have an exponential number of states at low enough temperatures \cite{bray1980metastable}, and it is not clear why the out of equilibrium dynamics should converge to equilibrium in this case. We remind the reader that the dynamics we are studying is obtained by taking the large $N$ limit first and thus activated processes between states are suppressed.

We want to put under a stringent test the above hypothesis and we find
convenient from the numerical point of view to test the so-called \textit{weak
  ergodicity breaking} property, stating that for any finite waiting time 
$t_w$ the correlation eventually decays to zero in the large time limit
\begin{equation}
\lim_{t\to\infty} C(t_w+t,t_w) = 0\quad\forall t_w\,.
\end{equation}
The physical meaning of this hypothesis is clear: in an aging system any
configuration reached at a finite time is eventually forgotten completely,
because the dynamics, although slower and slower, keeps wandering in a large
part of the configurations space. A direct consequence of the weak ergodicity
breaking is that an aging system does not keep memory of what it did at any
finite time, and the ``finite time regime'' can be eventually integrated out
(under a further hypothesis called weak long term memory: the long-time
dynamics wipes out the response to any small initial external perturbation), leading to an
asymptotic dynamics that is actually decoupled from the finite times
regime. This is a key feature that allows to get to the CK solution.

However, looking retrospectively at the literature of the times when the CK solution was
derived, one finds that the numerical and experimental evidence were far from definite.
While experiments do not probe a mean-field model, the numerics showed that the correlation $C(t_w+t,t_w)$ decays with time $t$ for each waiting time $t_w$
considered, but the range of correlations explored was very limited, with
correlations always relatively large: $C(t_w+t,t_w)\gtrsim 0.05$ \cite{baldassarri1998numerical,marinari1998}
More recently the Janus collaboration succeeded to probe much larger time
scales and much smaller correlations \cite{belletti2009depth}, but only studying finite dimensional spin
glasses for which the mean field solution is not clear to apply.

In the present work we report the results of an unprecedented numerical
effort in the study of the out of equilibrium dynamics in mean field spin
glasses to check whether some of the hypothesis at the basis of currently
available solutions are valid or not.

We started our study by simulating directly the SK model and found some
numerical evidence contrasting with the weak ergodicity breaking assumption
(these results are presented in detail in the SI). However we soon realized that such
a model would prevent us from reaching sizes and times scales large enough to
support any solid statement. Indeed the simulation time for such a model scales
quadratically with the system size $N$ and rapidly makes the simulation
unfeasible.

We then resorted to a spin glass model defined on a sparse graph
(requiring simulation times scaling linearly with the system size), 
similar to
the Viana-Bray (VB) model:~\cite{viana1985}
\begin{equation}
H_\text{\tiny VB} = -\sum_{(ij)\in E} J_{ij} s_i s_j\,,
\end{equation}
where the edge set $E$ defines the interaction graph and the couplings do not
need to be rescaled with the system size (as in realistic models).  In order
to reduce the effects of the spatial heterogeneity in the $H_\text{\tiny VB}$ Hamiltonian,
we simulate the model on a Random Regular Graph (RRG) of fixed
degree 4 and we use couplings of fixed modulus, i.e.\ $J_{ij}=\pm 1$, with
equal probability.  We measure self-averaging quantities, and for the very large sizes that we simulate there is no
visible dependence on the specific random graph or couplings realization.

One may question whether the out-of-equilibrium dynamics of the VB model is
equivalent to the SK case. The common belief about their equivalence is based on the 
observation that they are both mean field approximations of the same spin glass
model. Notice also that the SK model is equivalent to the large degree limit
of the VB model (once couplings or temperatures are properly rescaled).
However one may also argue instead that the SK and the VB models have a key
difference: only in the former the couplings become very weak in the
thermodynamic limit and this may produce visible differences in their out-of-equilibrium dynamics.
We do not have strong arguments against this point of
view and the only conclusive answer is to study both as we do in the present
paper and to compare the results.  Nonetheless let us argue that if the
out-of-equilibrium dynamics of the SK and VB were really asymptotically
different 
this would have important implications, and the VB model should be preferred to
the SK model as a mean field approximation to realistic spin glasses.  This
is one additional reason that convinced us to make such an important numerical
effort in understanding the large times out of equilibrium dynamics of the VB
model.

\section*{Results and discussion}

We have simulated the VB model on a RRG of degree 4 at temperature $T=0.8T_c$, where $T_c=1/\atanh(1/\sqrt{3})$,
starting from a random initial configuration. We have measured the correlation function $C(t_w+t,t_w)$
for $t_w=2^2,2^4,2^6,2^8,2^{10},2^{12}$ and very large times $t$.
Statistical errors have been strongly reduced averaging over a huge number of samples (see the Method section and the
SI for more details).

\begin{figure}[t]
\centering
\includegraphics[width=0.9\columnwidth]{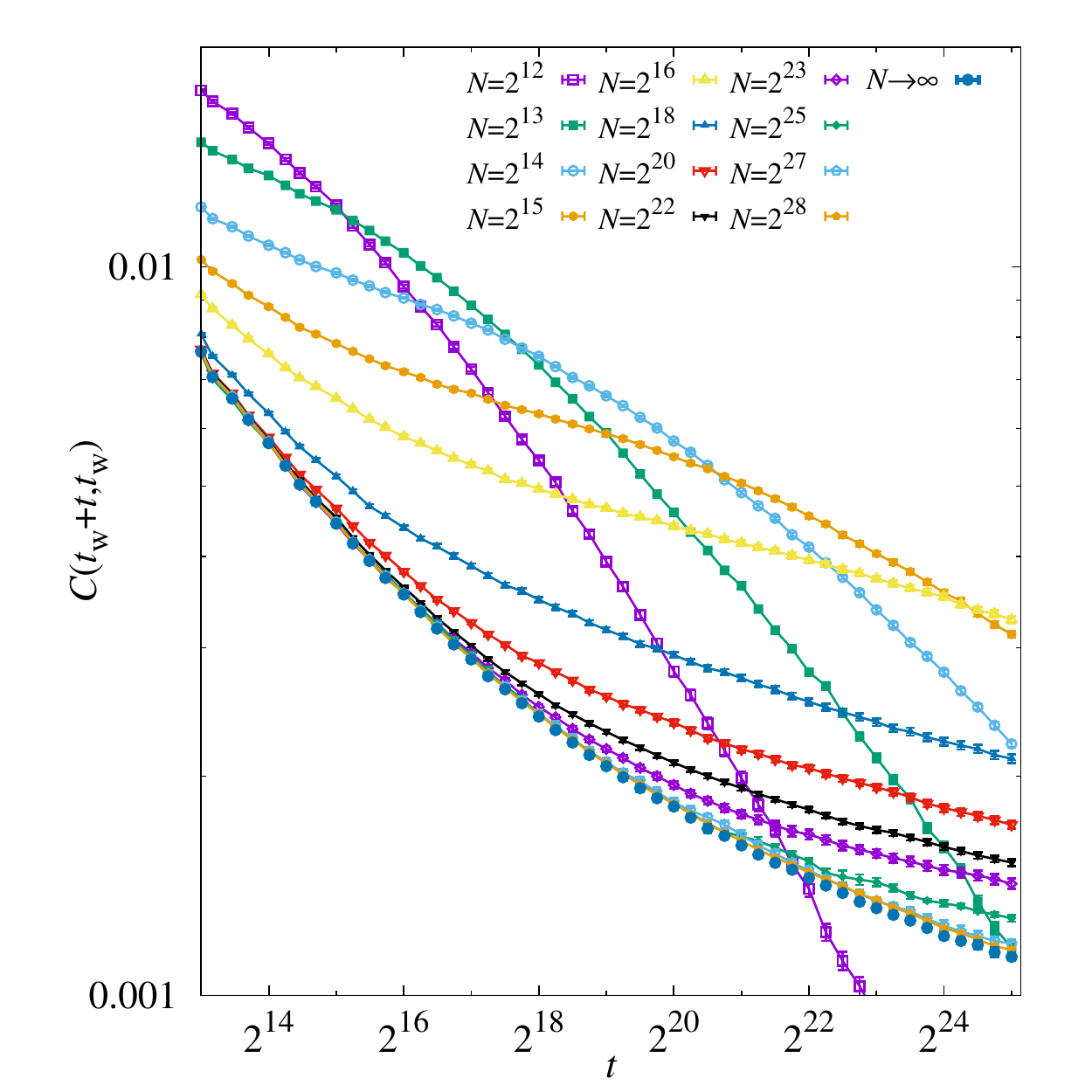}
\caption{Decay of the two-times correlation function $C(t_w+t,t_w)$ with
  $t_w=4$ in a spin glass defined on a random regular graph of degree 4,
  initialized in a random configuration and evolved at $T=0.8 T_c$.
  We show data for many different system sizes (lines are only a guide to the eye),
  and the extrapolation to infinite size for every value of $t$ (represented
  by the series of full symbols at the bottom).
  It is worth noticing that we are probing a regime of very large times and
  extremely small correlations, $C(t_w+t,t_w)\ll \qea$, never reached before
  in simulations of mean field models. The upward curvature in the thermodynamic
  limit is a strong indication against a power law decay to zero correlation,
  i.e.\ against the weak ergodicity breaking scenario.}
\label{fig:1}
\end{figure}

We show in Figure \ref{fig:1} the decay of the correlation function
$C(t_w+t,t_w)$ as a function of $t$ for $t_w=4$ and a number of different system sizes. The
reader should appreciate that we are working in the regime
$C(t,t')\ll\qea\simeq 0.285$ which has never been reached before in any study of the
out-of-equilibrium dynamics of mean field spin glass models
(the approximate value for $\qea$ is obtained via the replica
symmetric cavity method).  The plot is in a
double logarithmic scale, so an upward curvature is a clear indication that
correlation is either decaying slower than a power law or not decaying to zero at all.

Let us start discussing equilibration effects. For small enough $N$ we expect the system to thermalize and the correlation function to decay to the equilibrium value $C=0$. The thermalization time $t_\text{eq}(N)$ is strongly dependent on the system size $N$ and in Figure~\ref{fig:1} it is signalled by the shoulder clearly visible in the data for $2^{12}\le N\le 2^{15}$ and partially in the data for $N=2^{16}$.
Willing to study the out of equilibrium regime we must impose times to be much smaller than $t_\text{eq}$.
For $N \ge 2^{18}$ such a
thermalization effect is absent for the times we are probing and we can safely
consider the data as representative of the out of equilibrium regime.

Figure \ref{fig:1} shows that finite size effects become apparent when
measuring very small correlations. Such finite size corrections are otherwise
negligibly small at the correlation scale $C\sim 0.1$ which has been the mostly probed one in the past. It is worth
noticing that the clear identification of these finite size effects has been
possible thanks to the very small uncertainties that we have reached by
averaging over a very large number of samples (see the SI for details
on simulation parameters) and over a geometrically growing time window: in practice the
measurement at time $t$ is the average over the time interval $[t/2^{1/4},t]$.

In order to reach any definite conclusion in the analysis of the out-of-equilibrium dynamics,
it is mandatory to take into account these finite size
effects accurately in the attempt of extrapolating correlation data to the
thermodynamic limit. Indeed, as shown in Figure \ref{fig:1},
off-equilibrium correlations measured in a system of smaller size decay
asymptotically slower than those measured in a larger system. Then, without
taking properly the thermodynamic limit, one can not argue too much about the
large time limit of correlation functions.  This is the reason why the data we
initially got for the SK model, although pointing to the same conclusion we
will draw from VB model simulations, were considered not conclusive.

We are interested in the limit of very large times taken \emph{after} the
thermodynamic limit. In this limit we are probing the aging dynamics where
activated processes do not play any role.
For each \emph{fixed time} we
extrapolate the data shown in Figure \ref{fig:1} to the thermodynamic limit,
following the procedure explained in the SI, and we get the curve shown with
label ``$N\to\infty$'' in Figure \ref{fig:1}.  It is evident that a strong
upward curvature remains in the thermodynamical limit, thus suggesting
that a power law decay to zero is very unlikely and would require an unnatural very small
value of the power law exponent.

Hereafter we concentrate only on the analysis of data already extrapolated to
the large $N$ limit. We aim at understanding what is the most likely behavior
of the thermodynamic correlation in the limit of large times. We are aware
that solely from numerical measurements taken at very large but finite times
we cannot make an unassailable statement and one could always claim that on
larger times the decay could change.  Notwithstanding we believe (and we
assume in our analysis) that for the very large times we have reached in our numerical simulations the
asymptotic behavior already set in. Thus the results of the analysis should be
stable if we change the time window over which the analysis is carried out
(always in a way such that $t\gg t_w$ and thus $C(t_w+t,t_w)\ll \qea$).

\begin{figure}[t]
\centering
\includegraphics[width=0.9\columnwidth]{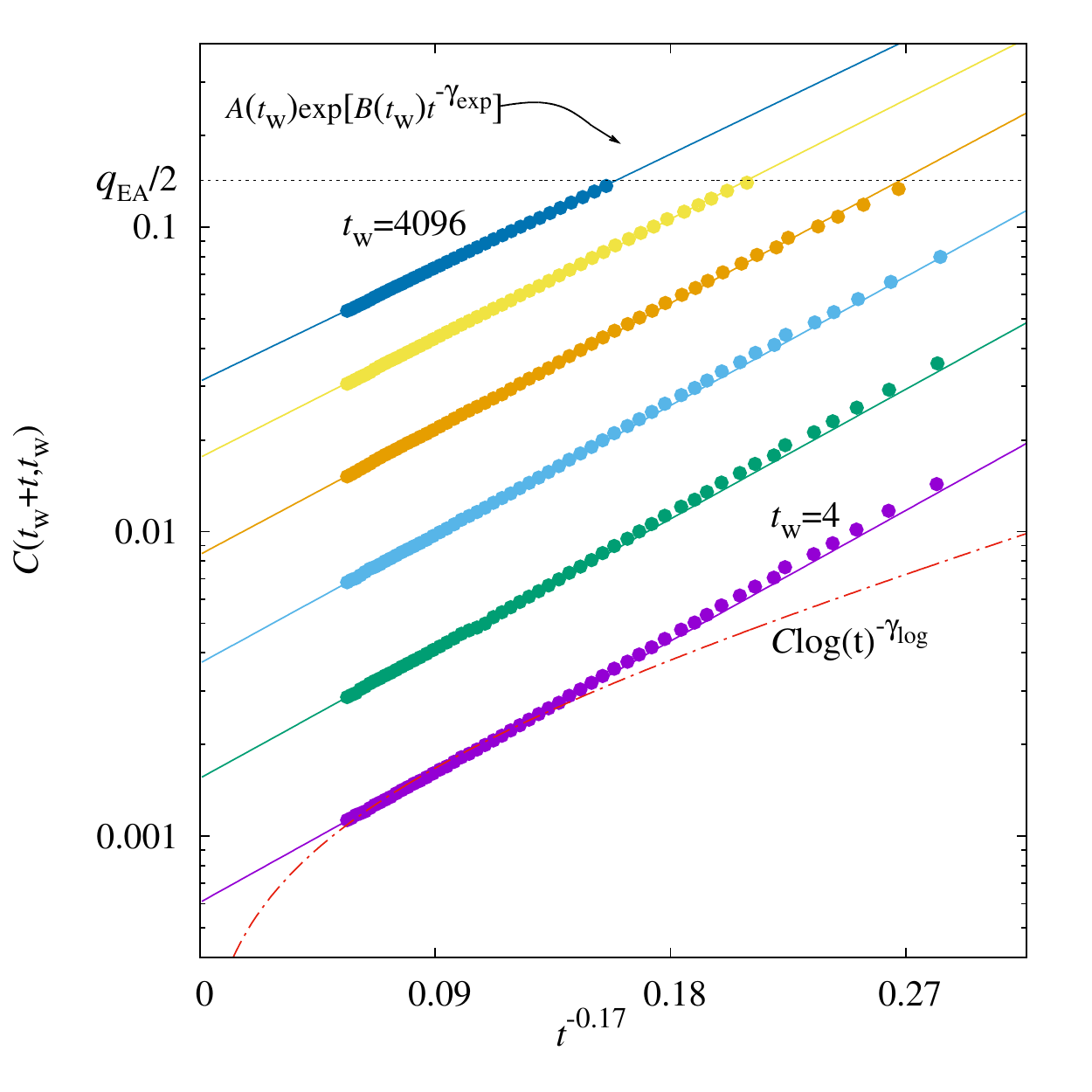}
\caption{The correlation functions $C(t_w+t,t_w)$ extrapolated in the thermodynamic
	limit as a function of time $t$, for all the waiting times $t_w$ studied.
  We report only data which are safely in the aging regime $C(t_w+t,t_w)<\qea/2$.
  Straight lines through data points are best fits to functional form in Eq.~[\ref{eq:fit}] on the time
  window $t\in[2^{17},2^{25}]$. The dashed curve is a best fit to a logarithmic
  decay extrapolating to zero correlation.}
\label{fig:2}
\end{figure}

In Figure \ref{fig:2} we present the results of the asymptotic analysis that
we find most stable and thus most likely, according to the above
prescription. The data for the correlations $C(t_w+t,t_w)$ extrapolated in the
large $N$ limit are shown as a function of an inverse power of the time for
different waiting times, ranging from $t_w=4$ to $t_w=4096$. We immediately
notice that all data follow a nice linear behavior in this scale; the only
data departing from the linear behavior are those at very short
times, that violate the condition $t_w\ll t$. The good agreement with the
linear behavior --- the lines are fits to data points
in $t\in[2^{17},2^{25}]$ --- implies that a fit to the form
\begin{equation}
C(t_w+t,t_w)=A(t_w)\exp\left[B(t_w)t^{-\gamma_{\text{exp}}}\right]
\label{eq:fit}
\end{equation}
is very stable upon changing the time window (mind the log scale on the y axis).
A joint fit to all $t_w$ data with a $t_w$-independent exponent gives the
value $\gamma_{\text{exp}}\simeq 0.17$, a weakly
$t_w$-dependent coefficient $B(t_w)$ and definitely a non-null extrapolations to
infinite time $A(t_w)=\lim_{t\to\infty}C(t_w+t,t_w)$.
For a quick reference we may call `exponential' the fit above, although the asymptotic
decay is like $A+A\,B\,t^{-\gamma_\text{exp}}$.

In order to test the null hypothesis, that is the weak ergodicity breaking scenario
where $\lim_{t\to\infty}C(t_w+t,t_w)=0$ for any finite $t_w$, we tried to fit
the data extrapolated in the large $N$ limit to a function compatible with this limit.
Given the upward curvature of the correlation function in a double
logarithmic scale (see Figure \ref{fig:1}) one may propose a very slow decay
according to an inverse power of $\log(t)$. It turns out that a fit to a logarithmic decay
$C(t_w+t,t_w)=C(t_w)\log{(t)^{-\gamma_{\text{log}}}}$, implying a null limiting correlation,
yields a value of the sum of squared residuals per degree of
freedom one order of magnitude larger than the fit in Eq.~[\ref{eq:fit}].
Moreover the values of the fitting exponents are also very strongly dependent on the
fitting window (see below and the SI).

The most stricking consequence of the analysis shown in Figure \ref{fig:2} is that, for finite $t_w$ values, 
in the $t\to\infty$ limit the correlation $C(t_w+t,t_w)$ does not decay to
zero. This is a very surprising result as it implies --- at variance with the
widely diffused common belief --- that an aging spin glass can asymptotically
remember, to some extent, the configurations it reached at \emph{finite
  times}. This positive long term correlation confutes the weak ergodicity
breaking assumptions and implies a much stronger ergodicity
breaking. The present result requires to rethink the asymptotic
solutions for the aging dynamics in mean-field spin glass models.

We show now some numerical evidence of why we consider the strong ergodicity breaking as the most
likely scenario. In order to test the stability of the fitting procedure
with respect to the choice of the time interval, we perform the analysis on
intervals $t\in[2^{n-k},2^n]$ with fixed $k=6$ and $n$ running on all the time
series. A fit to the function 
$C(t_w+t,t_w) =C(t_w) \log(t)^{-\gamma_\text{log}(t_w)}$ with $t_w$-dependent
parameters returns values of
$\gamma_\text{log}$ that strongly depend on $n$,
i.e.,\ on the position of the fitting window (see SI for details). Thus, fit results are very dependent
on the time $t$, implying strong corrections to the asymptotic
scaling. Again we can not completely exclude this scenario, but it is very
unlikely under the hypothesis that corrections to scaling are weak at the very
large times we reached at the end of our simulations.

The asymptotic scaling for the correlation decay which has been mostly used
until now is an inverse power law of time. Thus we have also fitted our data in the
time window $t\in[2^{n-k},2^n]$ according to low order polynomials in
$t^{-\gamma}$,
\begin{equation}
P_M(t_w+t,t)= A_M(t_w) + \sum_{m=1}^M D_M^{(m)}(t_w)\,t^{-m\gamma_{M}(t_w)}\;,
\label{eq:pol}
\end{equation}
with exponents and coefficients depending on both $t_w$ and $M$.

Please note that for $D_M^{(m)}(t_w)=A_{M}(t_w)B_{M}(t_w)^m$ the polynomial in Eq.~[\ref{eq:pol}]
is nothing but the $M$-th order Taylor expansion of the function in Eq.~[\ref{eq:fit}]
with $A_M(t_w)=A(t_w)$ and $B_M(t_w)=B(t_w)$.
It turns out that the identification $D=A\,B^m$ is necessary in
order to obtain numerically stable results. Moreover assuming such a relation we reduce the number of free
parameters and improve correlations among them.
At large times, where corrections to the asymptotic scaling should be
negligible, the low order polynomials should yield
results in agreement with those of the analysis made with the form in Eq.~[\ref{eq:fit}].
Any deviation would give a measure of the systematic error introduced
by ignoring some corrections terms in the asymptotic behavior.

\begin{figure}[t]
\centering
\includegraphics[width=0.8\columnwidth]{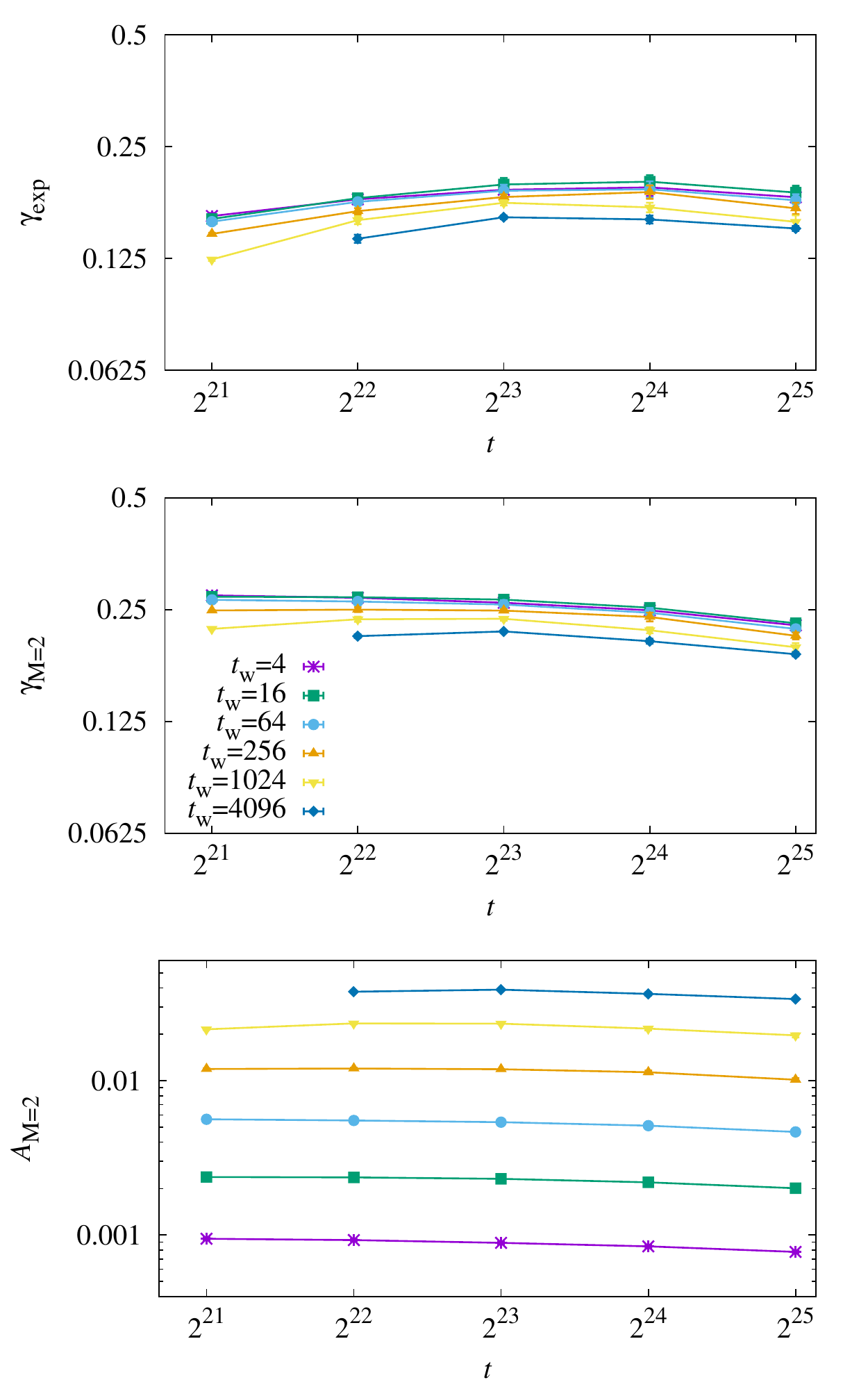}
\caption{The results of fitting $C(t_w+t,t_w)$ to Eq.~[\ref{eq:fit}] (upper panel) and to Eq.~[\ref{eq:pol}] with $M=2$ and assuming $D_M^{(m)}=A_{M}\,B_{M}^m$ (mid and lower panel). In all panels the time on the abscissa $t=2^n$ is the upper limit of the fitting range $t\in[2^{n-6},2^n]$. We notice that the results of these fitting procedure are very stable, i.e.\ vary little moving the fitting time window, at variance to fits assuming weak ergodicity breaking.}
\label{fig:3}
\end{figure}

We report in Fig.~\ref{fig:3} the best fitting parameters according to fitting functions in Eq.~[\ref{eq:fit}] and in Eq.~[\ref{eq:pol}] with $M=2$ and assuming $D_M^{(m)}=A_{M}\,B_{M}^m$. Results are shown as a function of the upper limit of the time window $[t/2^6,t]$ where the fit is performed. It is clear that the resulting best fit parameters are very stable, i.e.\ weakly dependent on the position of the fitting window, and this is a strong indication that we are probing the asymptotic regime with a functional form suffering only tiny finite time corrections.
We also notice that the two estimates of the decay exponent $\gamma_\text{exp}$ and $\gamma_{M=2}$, shown in the upper and mid panels, become compatible at large times. The asymptotic value for the correlation function $C(\infty,t_w)$, shown in the lower panel, is very stable too and clearly different from zero.

Analogous fits to any functional form assuming weak ergodicity breaking, that is $C(\infty,t_w)=0$, return best fitting parameters strongly dependent on the position of the fitting window, and a sum of squared residuals per degree of freedom\footnote{Since we are dealing with strongly correlated data, constructing a proper $\chi^2$ estimator is a challenging task, but the sum of the squared residuals can give anyhow an indication of the relative goodness of different interpolating functions.} which is typically one 
order of magnitude larger than the fits discussed above and illustrated in Fig.~\ref{fig:3}.

\begin{figure}[t]
\centering
\includegraphics[width=0.9\columnwidth]{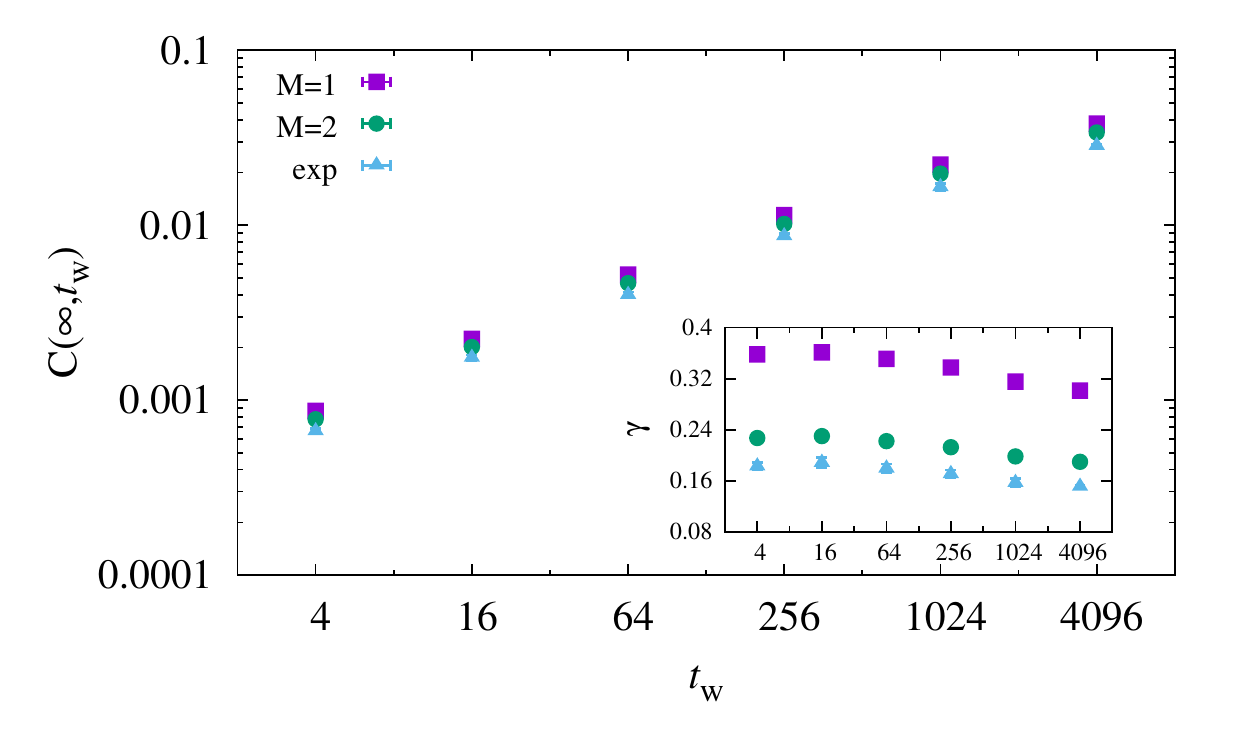}
\caption{The infinite time limit $C(\infty,t_w)$ estimated via the most stable fits,
	that is the one in Eq.~[\ref{eq:pol}] with $M=1$ and $M=2$, and the one in Eq.~[\ref{eq:fit}].
  The fitting range is $t\in[2^{17},2^{25}]$. Inset: the best estimate for the decay exponent.}
\label{fig:4}
\end{figure}

In conclusion the most likely scenario, which is fully supported by our data, is
the one where the limiting value for the correlation function $C(\infty,t_w)$
is strictly positive for any waiting time $t_w$. In Figure~\ref{fig:4} we
report the estimates of $C(\infty,t_w)=A_M(t_w)$ obtained from the last fitting window, $t\in[2^{17},2^{25}]$,
via the polynomial in Eq.~[\ref{eq:pol}] with $M=1$ and $M=2$, together 
with the values of $A(t_w)$ obtained via the fit to Eq.~[\ref{eq:fit}].
The three estimates are compatible within errors.
In the inset of Figure~\ref{fig:4} we report the best estimates for the decay exponent obtained from the same fits.
The exponent is weakly dependent on $t_w$, and systematic
errors are more evident, indicating that we are far from a regime in which the
decay to the residual correlation can be described by a single power
law. Notwithstanding this, different models with different decay exponents
agree both qualitatively and quantitatively with a non-null value for the
asymptotic correlation.

Computing the value of $C(\infty,t_w)$ in the large $t_w$ limit is out of
scope with present data and would require new and longer simulations with
larger $t_w$ values. However, given that we are working in the aging regime
under the condition $C(t_w+t,t_w)<\qea$, it is easy to get a conservative
upper bound to that limiting correlation, that is $\lim_{t_w \to \infty}
C(\infty,t_w) < \qea$.  Moreover, noticing that the plot in Figure \ref{fig:4}
is in double logarithmic scale and that we still do not see any downward
curvature, even if correlation values are not far from $\qea$, we may
conjecture that the upper bound is saturated, that is
\begin{equation}
\lim_{t_w \to \infty} C(\infty,t_w) = \qea\,.
\end{equation}
The validity of the above conjecture would lead to the unexpected scenario
where the out-of-equilibrium relaxation asymptotically gets trapped in an
equilibrium state, which is randomly chosen depending on the initial condition and the
dynamics at finite times.

The physical picture that emerges from the above strong ergodicity breaking scenario corresponds to a
system that, while relaxing in a complex energy landscape, remains confined in
regions of the configurations space becoming smaller and smaller during the
evolution. Whether this is a single state, a finite set of states or a
marginal manifold extending over just a finite fraction of the configurations
space is not possible to deduce from our data and further studies will be
needed.  Nonetheless if this strong ergodicity breaking scenario turns out to be the correct one (as our
data strongly suggest) we have to abandon the physical idea of aging as a
dynamical process exploring a marginal manifold extending all over the
configurations space.  The latter scenario can be still perfectly valid for
models defined on finite dimensional topologies (e.g.\ regular lattices)
because in that case barriers are not diverging exponentially with the system
size and so it is less likely to have a confining potential for the out-of-equilibrium
dynamics on finite timescales.

Recently the study of the out-of-equilibrium dynamics in a different mean
field spin glass model, namely the spherical mixed $p$-spin, has shown --- via
analytical solutions --- a similar phenomenon \cite{folena2019memories}: depending on the initial
condition the asymptotic aging dynamics may take place in a restricted part of
the configurations space, and the correlation with configurations at finite
times remain strictly positive.  This result corroborates those presented in
the present work and strongly suggests that, in mean field spin glasses, the most
general off-equilibrium relaxation is not the one we had in mind until now (a
slow and unbounded wandering in the entire configurational space), but a slow
evolution in a confined subspace, determined by the initial condition and the
early times dynamics.

In conclusion we have put under a severe test one of the most widely assumed hypothesis in the aging dynamics of mean-field glassy models, namely the weak ergodicity breaking scenario.
Our results are clearly in favour of a \emph{strong ergodicity breaking scenario}, where the two-times correlation function does not decay to zero in the limit of large times.
We have been able to achieve such unexpected result, thanks to (i) the use of sparse mean-field spin glass models, (ii) a new careful analysis taking care of both finite size and finite time effects and (iii) an extraordinary numerical effort based on very optimized codes running on latest generation GPUs.

It is fun to notice that the number of spin flips we performed on the largest simulated systems is of the order of $2^{64}$, the same number of wheat grains asked by Sessa, the inventor of chess, to sell his invention. Such an incredibly large number, that determined the destiny of Sessa, allows now to uncover unexpected physics!

\section*{Methods}

We simulated many samples of the VB model with sizes in the range
$2^{10}\le N\le 2^{28}$ for times up to $2^{25}$ Monte Carlo sweeps (MCS). We
report all details and parameters of the simulations in the SI.
Every simulation starts from a random initial configuration and evolves using
the Metropolis algorithm at temperature $T=0.8T_c=0.8/\atanh{\left(1/\sqrt{3}\right)}\simeq0.42$ \cite{mezard1987mean}.
This temperature is a good trade-off because it is not too low (and thus the
evolution is not too slow), while being in the low temperature phase and thus having an
Edwards-Anderson order parameter $\qea$ sensibly different from zero. We
remind that the aging dynamics takes place in the large times limit only under
the condition $C(t,t')<\qea$. The thermodynamical properties of typical samples of
the VB model can be computed via the cavity method: although for $T<T_c$ the
exact solution would require to break the replica symmetry in a continuous
way, we can get a reasonable approximation to the value of $q_\text{\tiny EA}$
via the replica symmetric solution providing $\qea(T=0.8T_c) \simeq 0.285$.

For simulating huge systems for long times it is necessary to resort to parallel processing. To that purpose we
  implemented the VB model on
  a random regular \textit{bipartite} graph (RRBG)  making possible to exploit the features of GPU accelerators
  despite of the very irregular memory access pattern (see SI).  We have checked on
intermediate sizes that results obtained on RRG and RRBG are statistically
equivalent (see results in the SI). The theoretical argument supporting the
statistical equivalence of the VB model defined on RRG and RRBG goes as
follows: RRG may have loops of any length, while RRBG only have even length
loops; since in the VB model with symmetrically distributed couplings (we use
$J_{ij}=\pm 1$) every loop can be frustrated with probability $1/2$
independently of its length, we do not expect any difference between the two
graph ensembles in the thermodynamic limit, where loops become long.
To speed up further the numerical simulations, we resorted to multispin
coding techniques where copies evolve in parallel on the same graph, 
but with different couplings and different initial
configurations.

Extrapolations to the thermodynamical limit is an important technical 
aspect of the present work: we dedicate to it a SI section.
The result presented here have been obtained by fitting to
a quadratic function $C(N=\infty)+AN^{-\nu}+BN^{-2\nu}$ with $\nu=2/3$.
The value of the exponent has been fixed according to well-known results in the literature~\cite{aspelmeier2008,boettcher2010,Lucibello2014}.

\acknow{
The research has been supported by the European Research Council under the EU Horizon
2020 research and innovation programme (grant No. 694925, G. Parisi).
}
\showacknow

\subsection*{References}

\bibliography{w}{}



\end{document}



\maketitle

\twocolumn

\section{The Sherrington-Kirkpatrick model}

Our interest in the asymptotic off-equilibrium dynamics stems from the
phenomenon of remanent magnetization of spin glasses. We recall that the
equations of dynamics of mean field models have been studied in detail in the
hypothesis of both weak ergodicity breaking and weak long term memory (see
main text)~\cite{cugliandolo1993analytical,cugliandolo1994out}. 
If a very strong constant magnetic field is applied to a spin glass,
the individual spins align along this magnetic field. If afterward
this magnetic field is switched off, the spin glass relaxes very
slowly towards a state of non-zero remanent magnetization, with some
excess internal energy relative to the internal energy at
equilibrium. This phenomenon has been the object of detailed
experimental studies~\cite{prejean1980,berton1982,omari1983}. On the theoretical side this
phenomenon has been studied in models, mostly by dynamical Monte Carlo
simulations.  If some numerical results have been
obtained~\cite{parisi1993remanent,parisi1997mean} for the finite dimensional Edwards
Anderson Ising model~\cite{ea1975}, most simulations have been
done~\cite{kinzel1986,henkel1987,parisi1993remanent,parisi1997mean,marinari1998} for the infinite range
Sherrington-Kirkpatrick model~\cite{sk1975,sk1978}.  These numerical results have
been obtained for small systems ($N\leq 1024$ in~\cite{kinzel1986,henkel1987},
$N\leq2048$ in~\cite{parisi1993remanent}, $N\leq 2016$ in~\cite{parisi1997mean}, but
$N\leq18432$ in~\cite{marinari1998}, where $N$ is the number of spins) using
either binary distributed quenched random couplings, or Gaussian
distributed quenched random couplings.

These pioneering simulations remain however inconclusive, and we decided
to revisit the case of the Sherrington-Kirkpatrick model with larger
systems (with up to $N=2^{15}=32768$ spins, quite a large number by
fully connected models standards), a larger time window, and a better
statistics.  We study the relaxation of the system following the Metropolis dynamics starting from a
configuration where all spins are aligned (for this model this is also
a disordered $T=\infty$ configuration). From a numerical point of view
this is a rare blessed situation for a Sherrington-Kirkpatrick
model simulation, where one is not restricted to a very small number of
spins, since there is neither the need to perform the lengthy
equilibration of the system (and to make sure that equilibrium is
indeed reached), nor the need to perform a satisfactory sampling of
the phase space in the subsequent measurement phase of the simulation.

We simulated the Sherrington-Kirkpatrick spin glass (model [1] in the main
text) with binary distributed quenched couplings, for temperatures
$T=0.4,\;0.5$ and $0.6$ ($T_c=1$ in this model).  The use of binary
distributed couplings allows a 
faster computer code, due to the multispin coding~\cite{friedberg1970test,jacobs1981multi}
technique to compute 
the force acting on a given spin (the computation of the force is the
bottleneck of the code). Away from $T=0$, the choice of binary
distributed couplings is believed to have little effect on the
physics.  Two clones are simulated in parallel.  The number of instances (i.e.\ disorder
coupling samples) is 
$N_{dis}=2^{24} / N $. This scaling of $N_{dis}$ with $N$ ensures
roughly $N$ independent statistical errors. The number of iterations
($N$ single spin updates, performed in a fixed order) is $2^{19}$. We 
lump together the data for the same value of 
$\lfloor\ln((t+t_w)/4)/(0.5\ln(2)) \rfloor$
where $\lfloor\cdots\rfloor$ is the floor function (\emph{tritone} logarithmic 
intervals for large times) and treat them as a data for the average value of 
$t+t_w$ inside the lump (e.g. we lump together the $t+t_w=\{4,5\}$ data, to be
treated as the data for $t+t_w=4.5$ in what follows).

We compute the two times correlation $C(t_w,t+t_w)$ as
\begin{equation}
C(t_w+t,t_w)=\frac{1}{N}\sum_{i=1}^N \langle\sigma_i(t_w+t) \sigma_i(t_w)\rangle\,
\label{eq:formula1}
\end{equation}
with a very small waiting time (we take $t_w=3$, we later realized
that a larger value would have given a better signal, see the main
text).  Angular brackets represent the average over the disorder and the two clones.

In this setting, the analysis of the large $t$ behavior of the data turns out
to be a quite difficult task,
due to a very limited range of values of $N$ together with the very
weak $t$ behavior of the correlation, and  the smallness of its $t=\infty$ limiting
value.

\begin{figure}[t]
\centering
\includegraphics[width=0.9\columnwidth]{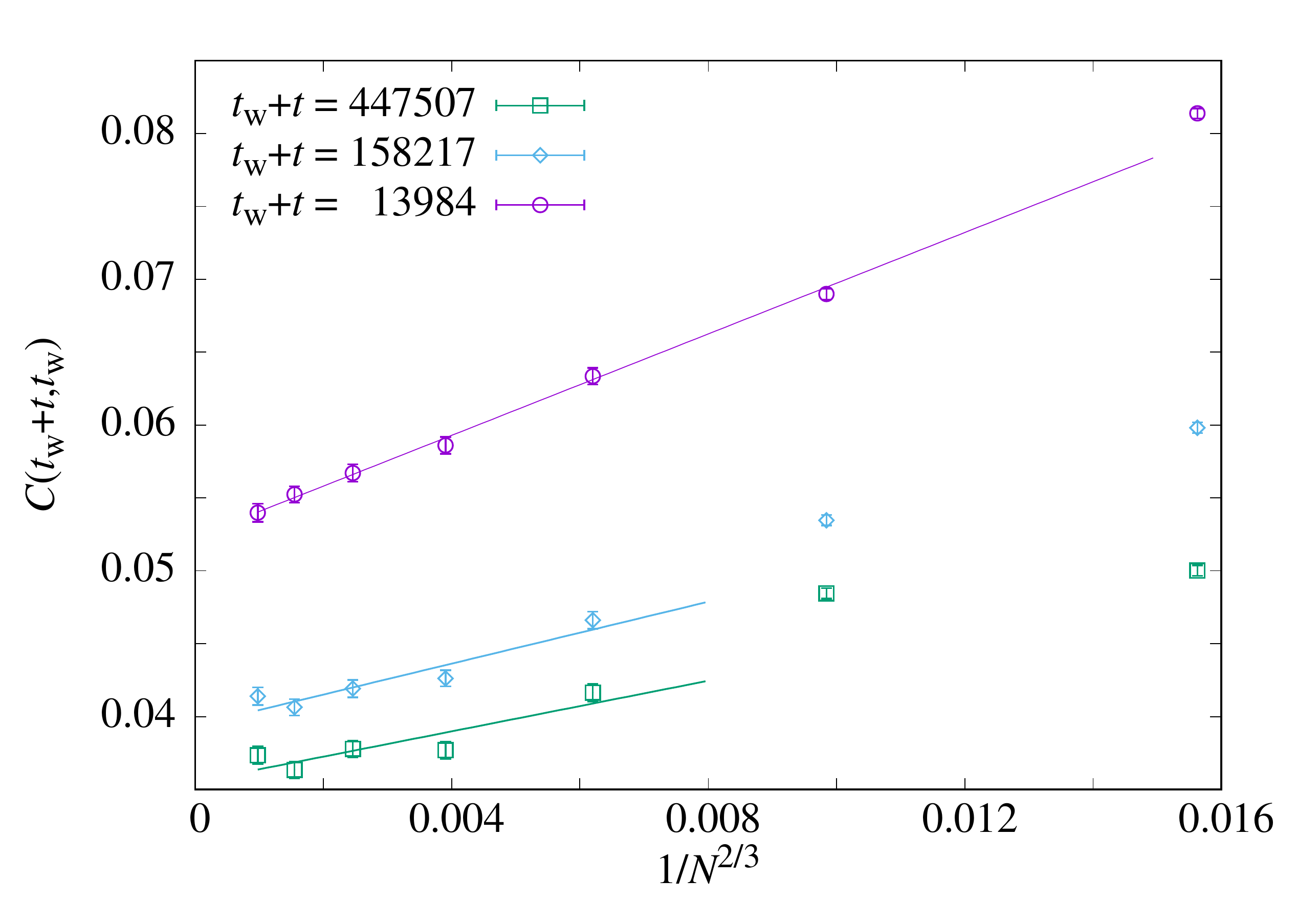}
\caption{Convergence of the two-times correlation function
  $C(t_w+t,t_w)$ as a function of $1/N^{2/3}$, for $T=0.5$ and  three
  representative values of $t$.}
\label{fig:un}
\end{figure}

We first extrapolate the data to the thermodynamic limit assuming a
leading $1/N^{2/3}$ behavior. This is done with a linear $\chi^2$ fit
of the data with $N\geq 2048$.  This extrapolation is shown in
figure~\ref{fig:un} for $T=0.5$ and three representative values of $t$.
The quality of the data deteriorates as $t$ grows, but remains
acceptable.  The slowness of the convergence to $N=\infty$ is a well
known feature of this model and clear subleading corrections are present
for small systems, whose sign turns out to depend on $T$.  We notice
that for all three temperatures the finite size corrections become
very small for the largest values of $t$.

We next try to determine the large $t$ behavior of (the infinite
volume limit) $C(t_w,t+t_w)$.  Our preferred fit of the data has the
form $A+B/t^{\gamma}$ with some very small $\gamma$ and  a non zero
$A$ but the value of $\gamma$ is unstable against details of the fitting
procedure, like the range of values of $N$ used in the $N\to\infty$
extrapolation, and the range of $t$ used to determine $\gamma$.  This
is to be expected since the range of values of $N$ and $t$ are limited
due to the full connectivity of the model. Note that a logarithmic decay
$C+D/\ln(t)^{\beta}$ is also possible.

\begin{figure}[t]
\centering
\includegraphics[width=0.9\columnwidth]{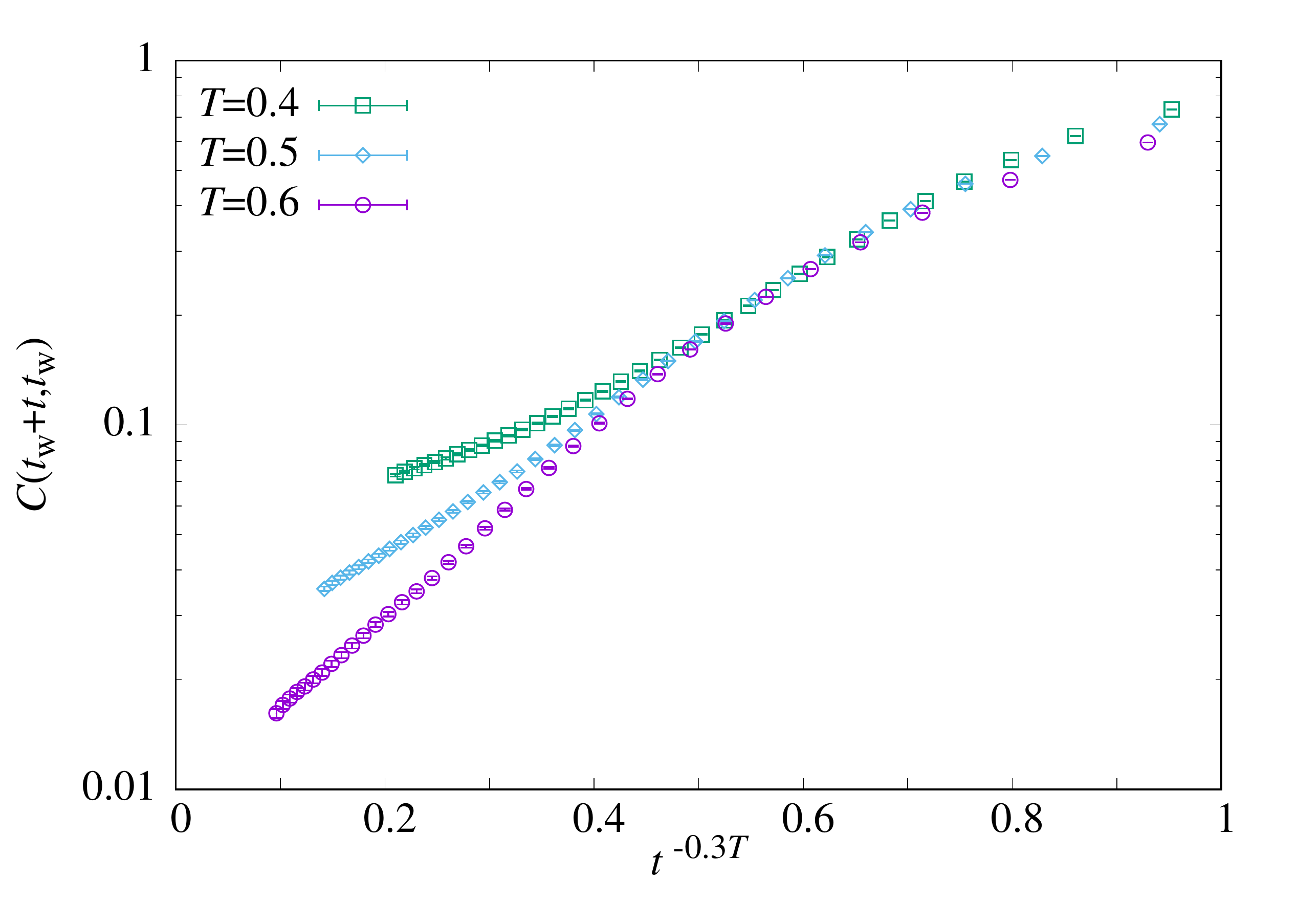}
\caption{$C(t_w+t,t_w)$ as a function of $1/t^{0.3 T}$ for $T=0.4,
  0.5$ and $0.6$.}
  \label{fig:deux}
\end{figure}

To make the case for a non zero limiting behavior we show in
Fig~\ref{fig:deux} a plot of $C(t_w,t+t_w)$ as a function of 
$1/t^{0.3T}$. There is a clear collapse of the data for an  intermediate range of
$t$. We interpret the collapse breakdown at larger values of $t$ as a
finite size effect.  This is however not a solid case and this
motivated us to perform the simulation of the sparse model presented
in the main paper. Such a system allows to simulate much larger systems,
for a longer time, with better statistics.

\section{Spin glass on a bipartite random regular graph: simulations and extrapolations}

We simulated instances (or \emph{samples}) of a spin glass with binary
coupling on a bipartite 
random regular graphs of degree $z=4$ (model~[5] in the main text) and number of
vertexes $N=2^{K}$ with $K=10-16,18,20,22,24,25,27,28$. We followed the
evolution of any instance 
starting from a completely random initial spin configuration applying the
classic single spin 
flip Metropolis algorithm, at fixed temperature $T=0.8T_c$,
up to time $t=2^{25}$, taking the update of the whole graph as unit of time. 
All samples had independent random coupling
configurations. In a straightforward multispin coding simulation the distinct bits of a data word represent the spins on a
single vertex of 64
independent samples, which must then share the same connectivity matrix. We
simulated a number of 
different graphs ranging between $3$ for  $N=2^{28}$ and $8192$ for $N=2^{10}-2^{14}$.
We performed simulations of samples with the smaller sizes
($K=10$ up to $16$ and $K=18$) on Intel CPUs with 64 bit data words (i.e.
64 independent random coupling configurations for each random graph
realization); simulations of larger sizes have been
accelerated by running a carefully coded program on Graphics Processing Units
with 64 bit data words (see Section~\ref{sec:GPU} below). Then the number of
simulated samples ranges between $192$ 
for $N=2^{28}$ to $524288$ for $N=2^{10}-2^{14}$ (see Table~\ref{tab:samples}).

\begin{figure}[t]
\centering
\includegraphics[width=0.9\columnwidth]{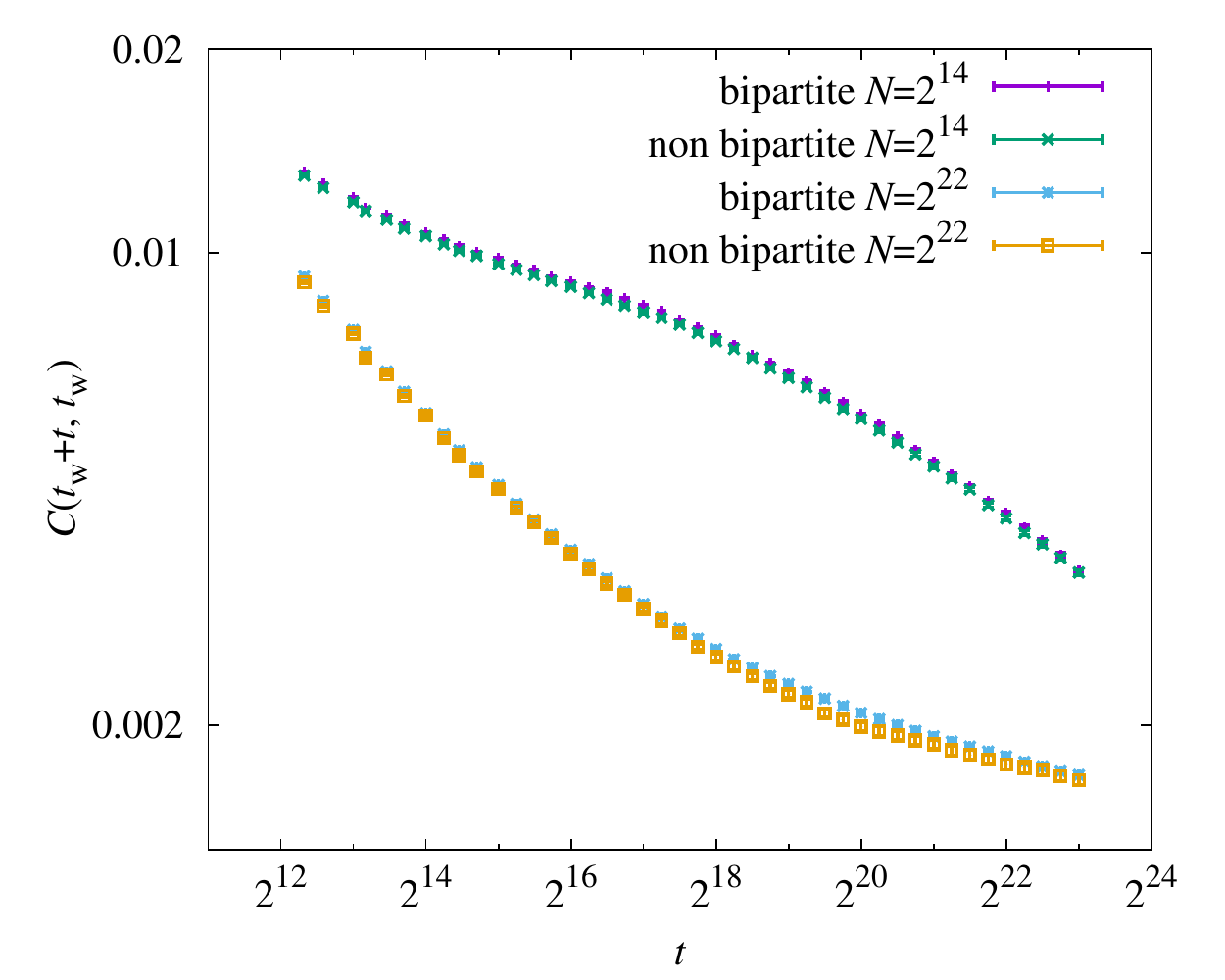}
\caption{Data of the autocorrelation function $C(t_w+t,t_w)$ for
$t_w=4$ and system sizes $N=2^{14}$ and $N=2^{22}$ for both bipartite and
  non-bipartite systems.}
\label{fig:bipnobip}
\end{figure}

Bipartition is mandatory in order to enforce parallelism in the Monte Carlo update (see
Section~\ref{sec:GPU}). This does not change the graph property of being
\emph{locally tree like} for large $N$ and then of approximating results on a Bethe lattice in
the $N\rightarrow\infty$ limit. Still, bipartition changes the distribution of lengths
of cycles and then possibly changes finite-size corrections to the behavior of the
system in the limit of the Bethe lattice \cite{lucibello2014}. We verified that bipartition is
statistically irrelevant to our results with the available 
numerical precision, on the smaller system sizes already, by comparison with
numerical simulations of the same model on non bipartite random regular graphs.  
Figure~\ref{fig:bipnobip} show data of the autocorrelation function $C(t_w+t,t_w)$ for
$t_w=4$ and system sizes $N=2^{14}$ and $N=2^{22}$ comparing results of averages over the
bipartite samples to averages over non bipartite samples. Even at the smaller
system sizes we do not observe any statistically significant difference in any time regime.

\subsection*{Extrapolation to large sizes and large times}
Hereafter, we describe firstly how we performed the limit to large sizes and
then to large times. 

We improved the signal to noise ratio by averaging the time series of
the measured $C(t_w+t,t_w)$ values, for each given waiting time $t_w$ and system
size $N$, over \emph{minor thirds} logarithmic intervals
$t_w+t\in\left[t_m,t_{m+1}\right]$, with $t_{m+1}/t_m=2^{1/4}$.

The extrapolation to large sizes at fixed times $t_w$ and $t$ is at the core
of our analysis. In order to take into account for finite size
corrections, we fitted data to both a linear function and a second order polynomial
in some power of the inverse system size $1/N^{\nu}$. 
We tried both $\nu=1$ and $\nu=2/3$, the latter being supported by the behavior of 
finite size corrections in the spin glass phase of mean field models~\cite{aspelmeier2008,boettcher2010,lucibello2014}.

\begin{figure}[t]
\centering
\includegraphics[width=0.9\columnwidth]{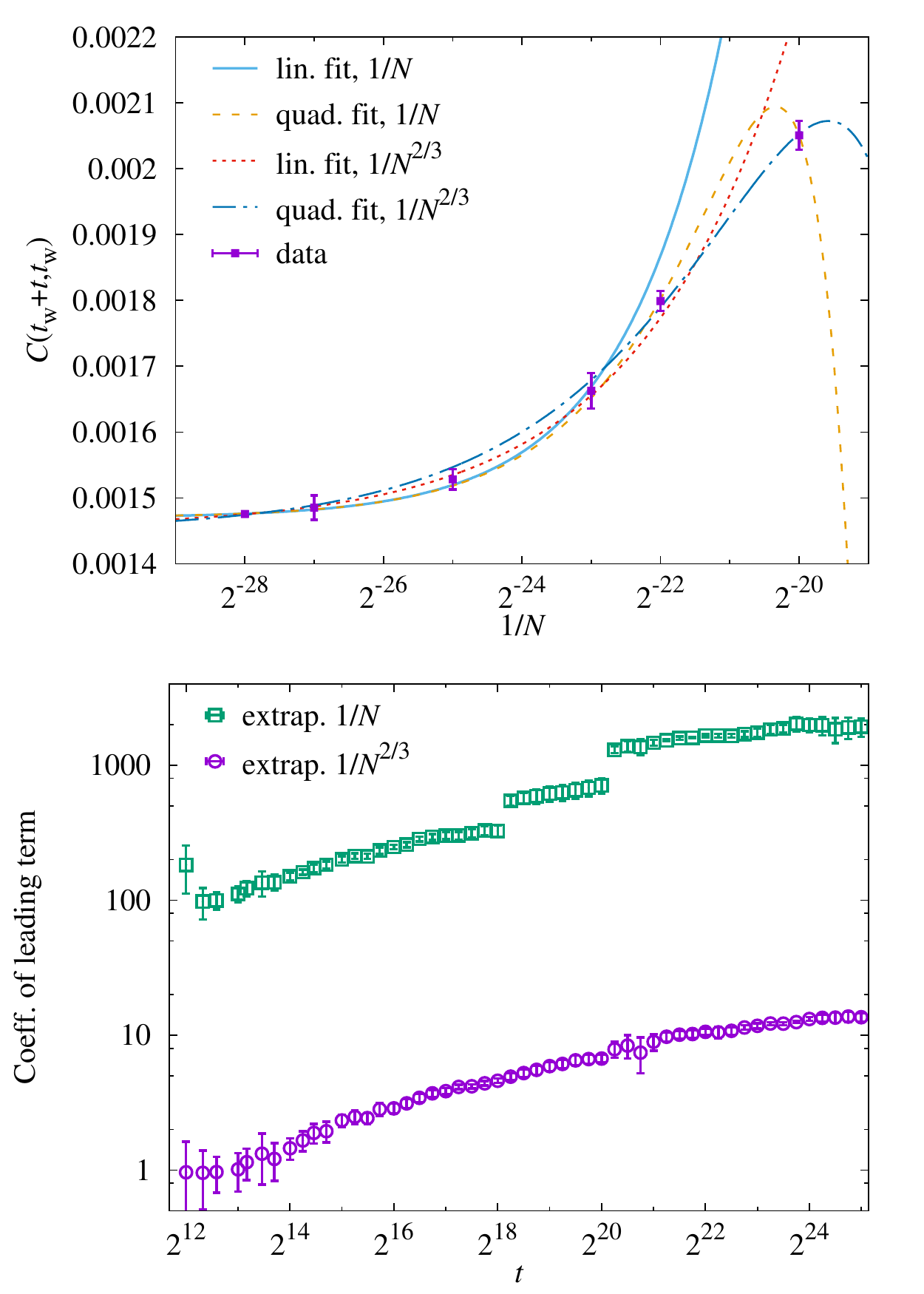}
\caption{Top panel: Data of the autocorrelation function for $t_w=4$,
  $t_w+t=2^{22}$, and superimposed adjusted functions (linear and quadratic
extrapolation to large sizes in both $N^{-1}$ and $N^{-2/3}$, see SI text.). 
Bottom panel: the adjusted values of the coefficients of the leading terms in the quadratic
extrapolations as a function of $t$ at $t_w=4$.}
\label{fig:fitpar}
\end{figure}

In the top panel of figure~\ref{fig:fitpar} we report a comparison of the four
functional forms in describing large-$N$ data for $t_w=4$, $t_w+t=2^{22}$.
Among the functional forms we employed, the quadratic fit  with
either $\nu=1$ or $\nu=2/3$ works equally well on almost all the $t_w$, $t$ range,
providing smooth extrapolations to infinite sizes.
By imposing $\nu=2/3$ we obtain a coefficient of the leading term of order 
$\sim O(1)$, whereas imposing $\nu=1$ results in a two orders of magnitude
larger value (see the bottom panel of figure~\ref{fig:fitpar}), suggesting
that the latter choice brings to a subestimation of finite size corrections.

We then performed extrapolations to large observation times (at fixed $t_w$) of
the values of the infinite size extrapolations of $C(t_w+t,t_w)$. After noting
the linear behavior of correlation data as a function of some power $\gamma$
in a log-linear plot (see Figure 2 and discussion in the main text), we tried
to fit data to an \emph{inverse Fr\'echet exponential}:
\begin{equation}
G(t)=A_\text{exp}(t_w)\exp{\left[B_\text{exp}(t_w)t^{-\gamma_\text{exp}}\right]}\;,
\label{eq:frechet}
\end{equation}
where the parameters $A_\text{exp}$, $B_\text{exp}$ and $\gamma_\text{exp}$ do
in principle depend on the waiting time. 
Results for the parameters are very stable when moving  the fitting window 
$t\in[2^{n-k},2^n]$ with $k=8$ by changing $n$ (see also Figure 3 and discussion in the
main text). Results are also stable when changing $k$ over a reasonable range
(smaller $k$ values imply noisy results, larger values return too few data points as
we have 12 octaves between our largest $t_w$ and the total time length of the simulation).
The exponent $\gamma_\text{exp}$ has a very weak dependence on the $t_w$.

In each window, and for any given value of $\gamma_\text{exp}$, we compute the
optimal values of $A_\text{exp}$, $B_\text{exp}$. We record the sum of
squared residuals as a function of $\gamma_\text{exp}$, the minimum of which is taken
as the optimum value of $\gamma_\text{exp}$. Since we deal with correlated
data, obtaining a $\chi^2$ estimator and uncertainties of parameters directly by the
fitting procedure is both involved and prone to noise. The sum of squared
residuals at the optimum $\gamma_\text{exp}$ value is still an useful metrics
in assessing relative accuracy of different models. 
To extract an estimation of uncertainties, we divide our database in three
subsamples and compute its maximum error.

\begin{figure}[t]
\centering
\includegraphics[width=0.9\columnwidth]{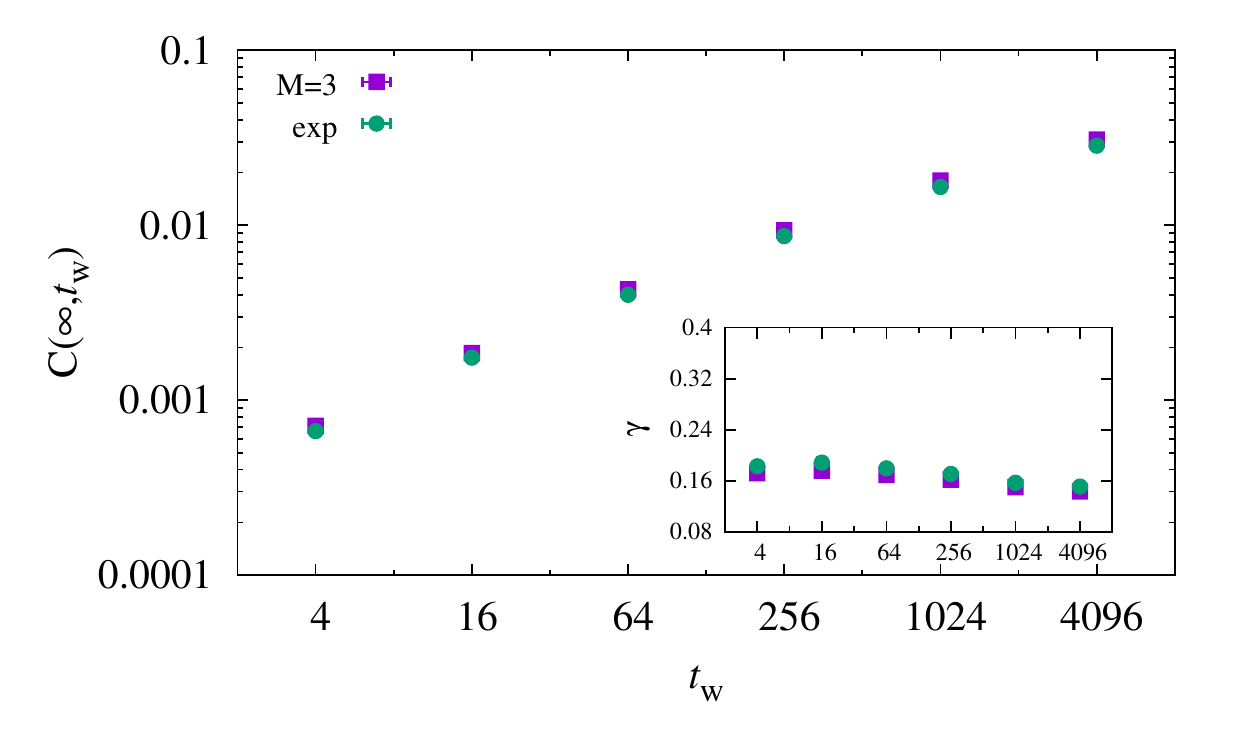}
\caption{The infinite time limit $C(\infty,t_w)$ estimated with the values of
  the constant terms in the $M=3$ polynomials Equation~\ref{eq:pol}
  and the coefficient $A_\text{exp}$ of the exponential function
  Equation~\ref{eq:frechet}, by fitting the models on the time window
  $t\in[2^{17},2^{25}]$.}
\label{fig:4_M3}
\end{figure}

One usually describes the large time off-equilibrium decay of the correlation
function with power laws. In order to test and quantify higher order
corrections to the asymptotic scaling we considered low order polynomials in
the variable $t^{-\gamma}$:
\begin{equation}
P_M(t_w+t,t)= A_M(t_w) + \sum_{m=1}^M A_MB_M^m(t_w)t^{-m\gamma_{M}(t_w)}\;,
\label{eq:pol} 
\end{equation}
With that choice of the parametrization, the $M$-th order polynomial is the
Taylor expansion to order $M$ of an exponential (Equation~\ref{eq:frechet}).
Such parametrization is also necessary to stabilize the fitting procedure
above described and reduce the parameter correlation, even for $M=1$. This provides a further confirmation
that the `exponential' decay works very well in all the large time regime.
The parameters $A_\text{exp}$ and $A_M$ in the last time window $[2^{17},2^{25}]$ 
are our estimation of the residual correlation.
The parameters  $A_M$, $B_M$ and $\gamma_M$ depend, in principle, on the waiting
time and on $M$. Also in this case the exponent $\gamma_M$ turns out having a
very weak $t_w$ dependence. The dependence of $\gamma_M$ on $M$ is significant
(see inset in Figure 4 in the main text), signalling strong subleading corrections
to the asymptotic, large time, behavior but the finite-time systematic error
on $A_M$ is comparable to the uncertainty (see also main panel in Figure 4
in the main text). $M=3$ data are already indistinguishable from results of
the exponential analysis, as shown in Figure~\ref{fig:4_M3}.

\begin{figure}[t]
\centering
\includegraphics[width=0.8\columnwidth]{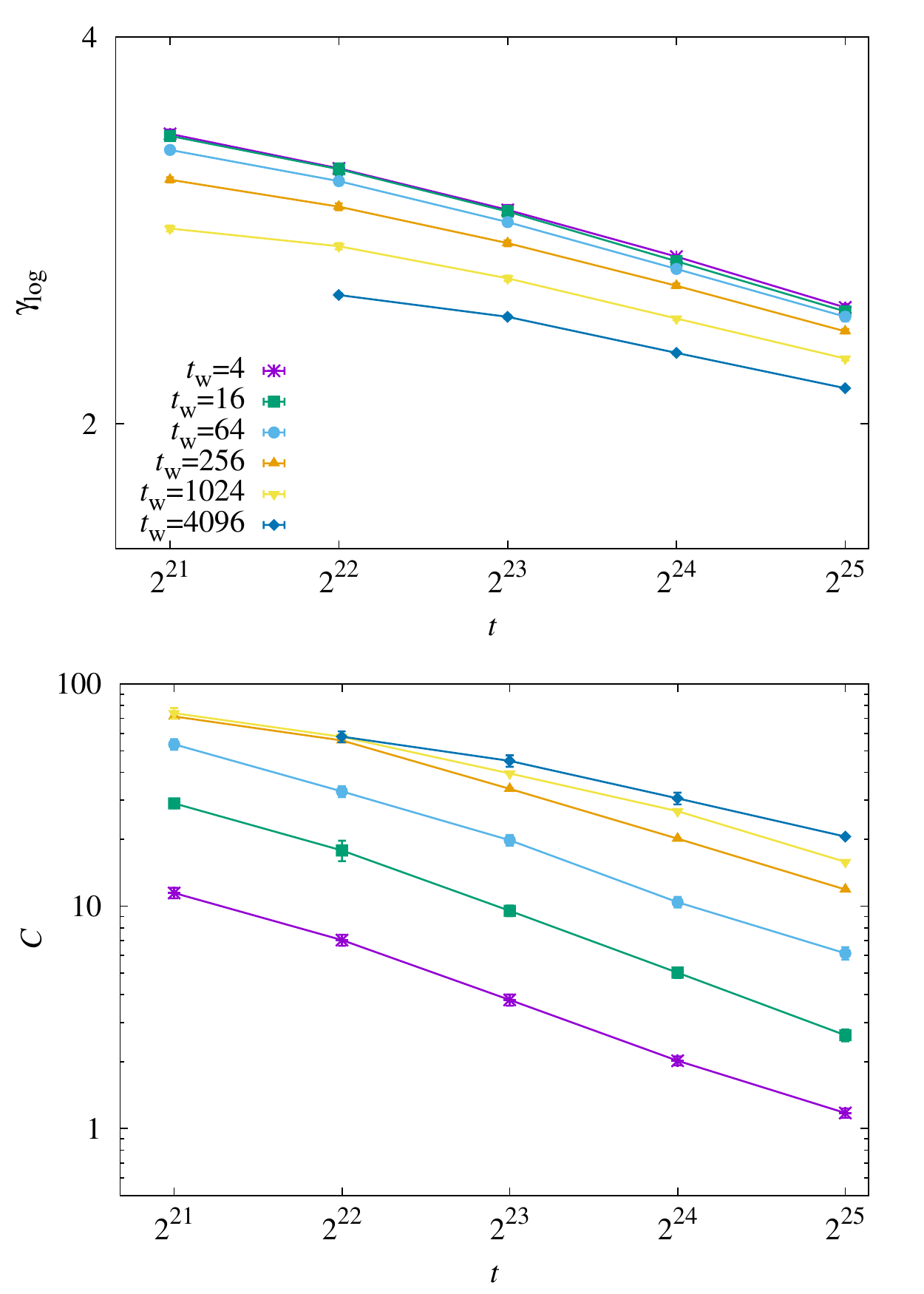}
\caption{Values of best fit parameters in the function
  $C [\log(t)]^{-\gamma_\text{log}}$ adjusted to $C(t,t_w)$
  data over a time window $[2^{n-k},2^{n}]$, with
$k=4$ and as a function of the right endpoint of the interval $t=2^n$. }
\label{fig:fitlog}
\end{figure}

A very slow logarithmic decay
\begin{equation}
L(t_w+t,t_w)=C(t_w)\left[\log(t)\right]^{-\gamma_\text{log}}
\end{equation}
can not in principle be ruled out. Yet we do not obtain encouraging results when
performing the fitting procedure on moving windows described above. As shown in
Figure~\ref{fig:fitlog}, results are much more dependent on the fitting windows,
and as a compensation for relatively quickly converging data,
the decay exponent takes much larger values than in the previous cases.

\section{Simulations of the Bethe Lattice on GPUs}
\label{sec:GPU}

Graphics Processing Units (GPU) provide a very good trade-off among
four crucial goals of any modern computing platform: performance,
price, power consumption and ease of software development. As a matter
of fact, although originally designed to accelerate graphics
rendering, GPU are, by now, widely used in many other applications of
high performance computing. The micro-architecture of a GPU is
massively parallel, with thousands of simple cores designed for
simultaneous, independent, not necessarily identical, computations on
multiple data inputs. For our work we resorted to NVIDIA GPUs
programmed by using the CUDA software framework. 

From a hardware standpoint, an NVIDIA GPU is an array of \emph{Streaming
  Multiprocessors} (SMs); each SM contains a certain number of CUDA
\emph{cores}. Each function executed on the GPU on behalf of the CPU
is called \emph{kernel}. To attain a significant fraction of the
theoretical peak performance, \emph{occupancy} (i.e., the fraction of
active computing elements at a given time) must be consistently kept
high, hence thousands of threads must be ready to be scheduled at any
time. Threads are executed in groups of 32 units called \emph{warps},
and performance is significantly improved if threads in the same warp
execute the same code with no divergence and access memory according
to patterns that privilege {\em threads} locality, {\em i.e.,} if
threads belonging to the same warp access consecutive memory locations
(memory {\em coalescing} in CUDA jargon). This is the most critical
issue in the simulation of the Ising Bethe lattice on a GPU since the
{\em neighbours} of each spin are, apparently, in {\em random}
positions with no regularity in the memory access pattern. As a matter
of fact, the only certainty is that each spin interact with a fixed
number (4) of other spins and that all of them are in the same half of
the lattice since the graph describing the connectivity among spins is
a bipartite graph. 

We highlight that the property of being a bipartite
graph is a fundamental requirement to update in parallel spins during
the simulation (it can be seen as a variation/extension of the classic
checkerboard decomposition used in the parallel update of spins having
nearest-neighbour interactions on regular lattices). As a consequence of this lack of
regularity, the memory access is not completely {\em coalesced} o,
more precisely, it is perfectly coalesced when reading the vector of
the interaction $J$ and writing the updated spins whereas it is not
coalesced reading the interacting spins. Despite of this limitation,
the performance of the GPU remains extremely good because the {\em
  kernel} in charge of the update uses just 32 registers so that up to
2048 threads per SM can be active (each SM has 65536 registers).
Having many threads active allows to alleviate the problems due to the
lack of memory coalescing. Warps alternate each other in the execution (while
waiting data coming from the memory) at zero cost, keeping the cores always
busy. The final result is that the update of a single spin take $\sim 15$
picoseconds on a P100 GPU. Actually that time can be further reduced by using
more than one GPU. 

Each half of the bipartite graph can be split between two or more GPUs with
each GPU updating a subset of spins. The idea is similar to what we presented
in Ref.~\cite{lulli2015highly} with the difference that in the present case each GPU must keep a full
copy of the spins due to the irregular connectivity. This means that much more
data must be exchanged among the GPUs (basically there is an all-to-all
communication exchanging the whole dataset of spins at each iteration). 
However, leveraging asynchronous memory copy operations and CUDA {\em streams}
it is possible to overlap communication and update of the spins achieving a
pretty good efficiency, at least with few GPUs ({\em e.g.,} $75\%$ on 4 GPU).
In the present study we were more interested in running many
simulations on a wide set of lattice sizes rather than running a long lasting 
single simulation on a lattice of fixed size so we did not use
extensively the multi-GPU variant of the code but for other case
studies it can significantly reduce the time-to-solution. A different
form of overlap that we, on the contrary, extensively used is that
between the update of the spins and the evaluation of the observables.
For taking measures we use the CPU that would be otherwise completely idle
during the simulation. When the measure must be carried out, we copy the
current spin configuration from the GPU to the CPU and then spawn a CPU thread
that executes the function which computes 
energy and correlations and saves the results in an output file. In this way
the GPU can continue the simulation while the CPU computes the
observables. 

The time required to copy the spin configuration from the GPU to
the CPU is approximately the same as the time required for the update of the
whole set of spins (obviously both depends on the size of the system but they
scale in the same way) whereas the time required to compute the correlations
is much higher since several spin configurations must be considered. So, in
the end, there is a significant time-saving in using the CPU to that purpose. 

\begin{table}[htb]
\centering
\caption{\normalfont Number of simulated samples of size
  $N=2^K$. $N_\text{samples}$ is the total number of independent random
  coupling configurations, while $N_\text{rrg}$ is the number of different
  instances of the random regular graph. On each graph instance we generate 64
  different random coupling configurations (which is the size of the 
  data word in the multipsin coding implementation).}
\begin{tabular}{crrrr}
\hline
\hline
 & \multicolumn{2}{c}{bipartite} & \multicolumn{2}{c}{non bipartite} \\
\hline
$K$ & $N_\text{samples}$  & $N_\text{bit}$ & $N_\text{samples}$ & $N_\text{rrg}$\\
\hline
$10-13$ &  $524288$ &  $8192$ &          &        \\
$14$    &  $524288$ &  $8192$ & $370176$ &  $5784$\\
$15$    &  $99264$  &  $1551$ &          &  \\
$16$    &  $75712$  &  $1183$ &          &  \\
$18$    &  $33280$  &  $520$  &          &  \\
$20$    &  $8192$   &  $256$  &          &  \\
$22$    &  $8256$   &  $129$  & $2560$   &  $40$ \\
$23$    &  $2048$   &  $32$   &          &  \\
$25$    &  $256$    &  $4$    &          &  \\
$27$    &  $192$     &  $3$    &          &  \\
$28$    &  $192$     &  $3$    &          &  \\
\hline
\hline
\end{tabular}
\label{tab:samples}
\end{table}

\FloatBarrier

\bibliography{w}{}